\documentclass[11pt,a4paper]{article}
\pdfoutput=1

\usepackage{jheppub}
\usepackage{latexsym}
\usepackage{revsymb}
\usepackage{multirow}
\usepackage{color}
\usepackage[usenames,dvipsnames,svgnames,table]{xcolor}

\usepackage{graphicx}
\usepackage{epsfig}  
\usepackage{epsf}    
\usepackage{dcolumn}
\usepackage{bm}
\usepackage{dcolumn}
\usepackage{textcomp}
\usepackage{float}
\usepackage{hypcap}
\usepackage[]{hyperref}
\usepackage{adjustbox}
\usepackage{multirow}

\usepackage{amsmath}
\usepackage{subfigure}
\usepackage{alphalph}
\usepackage{morefloats}
\usepackage{textcomp}
\usepackage[utf8x]{inputenc}
\usepackage{float}
\restylefloat{table}

\newcommand{\be}{\begin{equation}}
\newcommand{\ee}{\end{equation}}
\newcommand{\ba}{\begin{eqnarray}}
\newcommand{\ea}{\end{eqnarray}}

\newcommand{\chisq}{\chi^2}

\title{Determining Neutrino Mass Ordering with ICAL, JUNO and T2HK} 

\author[a,b]{Deepak Raikwal,}
\author[c,d]{Sandhya Choubey,}
\author[e,f]{Monojit Ghosh}

\affiliation[a]{Harish-Chandra Research Institute,  A CI of Homi Bhabha National Institute, Chhatnag Road, Jhunsi, Prayagraj - 211019}
\affiliation[b]{Homi Bhabha National Institute, Anushakti Nagar, Mumbai 400094, India}
\affiliation[c]{Department of Physics, School of Engineering Sciences, KTH Royal Institute of Technology,\\ AlbaNova University Center, Roslagstullsbacken 21, SE--106 91 Stockholm, Sweden }
\affiliation[d]{The Oskar Klein Centre, AlbaNova University Center, Roslagstullsbacken 21,\\ SE--106 91 Stockholm, Sweden}
\affiliation[e]{School of Physics, University of Hyderabad, Hyderabad - 500046, India}
\affiliation[f]{Center of Excellence for Advanced Materials and Sensing Devices, Ruder Bo\v{s}kovi\'c Institute, 10000 Zagreb, Croatia}

\emailAdd{deepakraikwal@hri.res.in}
\emailAdd{choubey@kth.se}
\emailAdd{monojit$\_$rfp@uohyd.ac.in}

\abstract{
In this paper we study the synergy among the future accelerator (T2HK), future atmospheric (ICAL) and future reactor (JUNO) neutrino experiments to determine the neutrino mass ordering. T2HK can measure the mass ordering only for favorable values of $\delta_{\rm CP}$, whereas the mass ordering sensitivity of JUNO is dependent on the energy resolution. Our results show that with a combination of T2HK, ICAL and JUNO one can have a mass ordering sensitivity of 7.2 $\sigma$ even for the unfavorable value of $\delta_{\rm CP} = 0^\circ$ for T2HK and most conservative value of JUNO energy resolution of 5$\%/\sqrt{E(MeV)}$.
The synergy mainly comes because different oscillation channels prefer different values of $|\Delta m_{31}^2|$ in the fit when the mass-ordering $\chi^2$ is minimized. In this context we also study: (i) effect of varying energy resolution of JUNO, (ii) the effect of longer run-time of ICAL,  (iii) effect of different true values of $\theta_{23}$ and (iv) effect of octant degeneracy in the determination of neutrino mass ordering.
}

\arxivnumber{}
\begin{document}
\maketitle

\section{Introduction}
\label{intro}

Neutrino oscillation in the standard three flavour frame-work is defined by three mixing angles: $\theta_{12}$, $\theta_{23}$ and $\theta_{13}$, two mass squared differences: $\Delta m^2_{21}$ and $\Delta m^2_{31}$ and one Dirac type CP phase $\delta_{\rm CP}$. At this moment one of the major unknowns in the standard three flavour paradigm is the neutrino mass hierarchy or the true ordering of the neutrino masses. From the solar neutrino experiments we already know that $m_2 > m_1$, which leads to $\Delta m^2_{21} = m_2^2 - m_1^2 > 0$. However, the sign of $\Delta m^2_{31} = m_3^2 - m_1^2$ is still unknown. It can be either greater than zero i.e., $m_3 > m_1$, giving rise to normal ordering (NO) of the neutrino masses or it can be less than zero i.e., $m_3 < m_1$, giving rise to inverted ordering (IO) of the neutrino masses~\cite{Gonzalez-Garcia:2021dve}. Amongst the experiments that are currently running and are somewhat sensitive to the neutrino mass ordering are the atmospheric based Super-Kamiokande~\cite{Super-Kamiokande:2017yvm}, and the  accelerator-based T2K~\cite{T2K:2021xwb} and NO$\nu$A~\cite{NOvA:2021nfi}. The combined analysis of the data obtained from these experiments shows a preference for normal ordering over inverted ordering~\cite{Esteban:2020cvm}. Apart from neutrino mass ordering, the other two unknowns in the three flavour scenarios are: the octant of the atmospheric mixing angle $\theta_{23}$ which can be either in the lower octant i.e., $\theta_{23} < 45^\circ$ or in the higher octant i.e., $\theta_{23} > 45^\circ$, and the phase $\delta_{\rm CP}$. Regarding the octant measurement of the current experiments, there is a mild preference towards the higher octant but the maximal value i.e., $\theta_{23} = 45^\circ$ is also allowed at $3 \sigma$. Regarding $\delta_{\rm CP}$, there is a mismatch between the values obtained from T2K and NO$\nu$A. T2K provides a best-fit value of $\delta_{\rm CP}$ around $-90^\circ$~\cite{con_talk_t2k} whereas the best-fit value from NO$\nu$A is around $\pm 180^\circ$~\cite{con_talk}.

The major proposed future experiments which are designed to establish this hint to a significant confidence level are: accelerator based T2HK~\cite{Abe:2016ero} and DUNE~\cite{Abi:2020evt}, atmospheric based ICAL detector of INO facility~\cite{ICAL:2015stm} (from now on we will refer to it as ICAL), ORCA~\cite{KM3Net:2016zxf} and PINGU~\cite{IceCube:2016xxt}, and reactor based JUNO~\cite{JUNO:2015zny}. In this paper we study the potential of the future experiments ICAL, JUNO and T2HK to determine the neutrino mass ordering. Though each of them individually have good chances of pinning down the neutrino mass ordering, they also have their limitations. In this work, we highlight those limitations and show how one can overcome those limitations by invoking synergistic combination of two or all three of them. In particular, we will probe the role of $|\Delta m^2_{31}|$ when we fit the simulated data to test neutrino mass ordering in these experiments. We will see that since different oscillation channels when fitted return different values of $|\Delta m^2_{31}|$, the overall $\chi^2$ obtained in a combined analysis of two or all three of these experiments returns a $\chi^2$ that is significantly larger than a simple sum of the individual $\chi^2$. 

Note that synergistic study between future atmospheric and future reactor experiments, future accelerator and future atmospheric experiments and future accelerator and future reactor experiments to determine neutrino mass ordering has been performed in the past. For example, Ref.~\cite{IceCube-Gen2:2019fet} studies the synergy between PINGU and JUNO, synergy between ORCA and JUNO is studied in Ref.~\cite{juno_orca}, Ref.~\cite{Blennow:2013vta} studies the synergy between reactor based Daya Bay II~\cite{wang} and PINGU, Ref \cite{Fukasawa:2016yue} studies the synergies among T2HK, DUNE and Hyper-Kamiokande, Ref.~\cite{Chakraborty:2019jlv} studies synergy between accelerator based ESSnuSB \cite{ESSnuSB:2021lre} and ICAL, and synergy in T2K II, NO$\nu$A II and JUNO can found in Ref.~\cite{Cao:2020ans}. The synergy in the combined beam and atmospheric data of DUNE is studied in Refs.~\cite{Barger:2013rha,Barger:2014dfa} and the same for ESSnuSB is studied in~\cite{Blennow:2019bvl}. Recently in Ref.~\cite{Choubey:2022gzv}, we have studied the synergistic effect of T2HK and JUNO to measure neutrino mass ordering.
However, to the best of our knowledge, a synergistic study among the accelerator, atmospheric and reactor experiments considering these particular set of experiments i.e., ICAL, JUNO and T2HK has not been performed in the past and this paper is the first of this kind. The mass ordering sensitivity of T2HK experiment is dependent on the values of $\delta_{\rm CP}$ and can have mass ordering sensitivity for only favourable values of $\delta_{\rm CP}$ i.e., $-180^\circ < \delta_{\rm CP} < 0^\circ$. For JUNO, the mass ordering sensitivity is dependent on the energy resolution whereas the mass ordering sensitivity of ICAL is hardly dependent on $\delta_{\rm CP}$. Therefore, in this work we will show that when these three experiments are combined, one can achieve mass ordering sensitivity at a significant confidence level irrespective of the energy resolution of JUNO or the value of $\delta_{\rm CP}$. The main synergy between the reactor experiment and the accelerator/atmospheric experiment come from their sensitivity to the parameter $\Delta m^2_{31}$. In addition we will also explore (i) effect of varying energy resolution of JUNO, (ii) the effect of longer run-time of ICAL,  (iii) effect of different true values of $\theta_{23}$ and (iv) effect of octant degeneracy in the determination of neutrino mass ordering.

The paper is organized as follows. In the next section we will briefly describe the experimental specification which we consider for our simulation. In Section \ref{res} we will present our detailed results. Finally in Section \ref{sum} we will summarize our results and conclude.

\section{Experimental setup and simulation details}
\label{spec}
In this section we will briefly describe the specifications of the experiments which we use in our simulation. For ICAL, we use the GEANT4-based geometry of ICAL detector with specification given in Ref.~\cite{ICAL:2015stm}. ICAL is a 50 kton iron  calorimeter with $\le$ 1.5 T magnetic field which makes it world's unique detector which can distinguish between charged current atmospheric neutrino and antineutrino events. We use the Honda atmospheric neutrino fluxes calculated for Theni site in India \cite{Honda:2015fha}. We use the GENIE event generator \cite{Andreopoulos:2009rq} developed for ICAL by the INO collaboration. We generate 1000 years data for ICAL and then normalize it to 10 years to reduce the Monte Carlo fluctuations. Neutrino oscillations are calculated numerically and incorporated via the re-weighting algorithm. The detector efficiency, muon energy and angle resolutions, the charge identification efficiency of the muons, and the hadron energy resolution provided via simulation work done by the INO collaboration are included \cite{muon-reso,hadron-resol}. We bin the simulated events in bins of muon energy ($E_{\mu}$), muon angle ($\cos\theta_{\mu}$) and hadron energy ($E_{\rm had}$).
The binning scheme used is shown in Table~\ref{tab:inobin} \cite{moon:2014,Mohan:2016gxm}. The analysis is performed by defining a $\chi^2$ as given in \cite{moon:2014}. For systematic errors we use (i) flux normalization error of 20$\%$ (ii) cross-section normalization error of 10$\%$ (iii) 5$\%$ uncertainty on the zenith angle dependence of the fluxes (iv) 5$\%$ energy-independent systematic uncertainty and an (v) energy dependent tilt factor. See \cite{anu-mh} for further details.

\begin{table}[h!]
\begin{center}
\begin{tabular}{ |c|c|c|c| } 
 \hline
 Observable & Range & Bin width & No. of bins\\ 
\hline
 $E^{\rm obs}_{\mu}$(GeV)(15 bins) & [0.5,4] & 0.5 & 7\\
 						 & [4,7]    & 1   & 3\\ 
						& [7,11]   & 4   & 1 \\
						& [11,12.5] & 1.5 & 1\\
						&[12.5,15] & 2.5&1\\
						&[15,25] & 5&2\\
$\cos(\theta^{\rm obs}_{\mu}$ (21 bins)       &   	[-1.0,-0.4]  &0.0.05&12\\
			&[-0.4,0.2]&0.10&2\\
						&[-0.2,1.0]&0.2&6\\
$E^{\rm obs}_{\rm had}$ (GeV)  (4bins)               &[0,2]&1&2\\
						&[2,4]&2&1\\
						&[4,15]&11&1\\

 \hline
\end{tabular}
\caption{Details of the three observable $E_{\mu}$, $\theta_{\mu}$ and $E_{\rm had}$ which are used in the analysis.}
\label{tab:inobin}
\end{center}
\end{table}

The experiments T2HK and JUNO are simulated using GLoBES \cite{Huber:2004ka,Huber:2007ji}. 
For T2HK, we follow the configuration as given in Ref.~\cite{Abe:2016ero}. We consider two water-Cerenkov detector tanks having fiducial volume of 187~kton each located at Kamioka which is 295 km from the neutrino source at J-PARC having a beam power of 1.3 MW with a total exposure of $27 \times 10^{21}$ protons on target, corresponding to 10~years of running. We have divided the total run-time into 5 years in neutrino mode and 5 years in anti-neutrino mode. For systematic errors, we have considered overall normalization error of 4.71\% (4.13\%) for the appearance (disappearance) channel in neutrino mode and 4.47\% (4.15\%) for the appearance (disappearance) channel in antineutrino mode \cite{Abe:2016ero}. The systematic error is the same for both signal and background. 

For JUNO, we consider the same configurations as given in Ref. \cite{JUNO:2015zny}. We consider a liquid scintillator detector having 20 kton fiducial mass located at a distance around  53 km from Yangjiang and Taishan nuclear power plants. We have considered the energy resolution of $3\%/\sqrt{E(MeV)}$ unless otherwise mentioned. In this analysis we consider all the reactor cores are located at the same distance from the detector. We use 200 same-size bins within the energy window of 1.8 MeV and 8.0 MeV. We have taken the backgrounds and systematic uncertainties as presented in \cite{JUNO:2015zny}. We consider the run-time to be 6 years.

Mass ordering sensitivity of an experiment is defined by its capability to exclude the wrong mass ordering. As the global data shows a preference towards normal mass ordering we will present our main results for normal ordering. For the estimation of the sensitivity, we use the Poisson log-likelihood:
\begin{equation}
 \chi^2_{{\rm stat}} = 2 \sum_{i=1}^n \bigg[ N^{{\rm IO}}_i - N^{{\rm NO}}_i - N^{{\rm NO}}_i \log\bigg(\frac{N^{{\rm IO}}_i}{N^{{\rm NO}}_i}\bigg) \bigg]\,,
 \label{eq:chi2}
\end{equation}
where $N^{{\rm IO}}$ is the number of events expected for the values of the oscillation parameters tested for, $N^{{\rm NO}}$ is the number of events  expected for the parameter values assumed to be realized in Nature (Asimov dataset) and $i$ is the number of energy bin.
The values of the oscillation parameters used in this analysis is given in Table \ref{Tab:Param}.
\begin{table}[h]
	\centering
	\setlength{\extrarowheight}{0.1cm}
	\begin{tabular}{|c|c|c|}
		\hline
		 Parameter & Best-fit value & $3 \sigma$ range \\
		\hline
		$\theta_{12}$ & $33.44^\circ$ & - \\
		$\theta_{13}$ & $8.57^\circ$ & - \\
		$\theta_{23}$ & $45^\circ$ & $40^\circ$ to $52^\circ$ \\
		$\delta_{\rm CP}$ & $-90^\circ/0^\circ$ & $-180^\circ$ to $180^\circ$  \\
		$\Delta m^2_{21}$ & $7.42\times10^{-5}$ eV$^2$ & -  \\
		$\Delta m^2_{31}$ &  $2.531\times10^{-3}$ eV$^2$ & (2.435 to 2.598) $\times 10^{-3}$ eV$^2$ \\
		\hline
	\end{tabular}
	\caption{The best-fit values and $3 \sigma$ ranges of the oscillation parameters used in our calculation.}
	\label{Tab:Param}
\end{table}
We have kept the parameters $\theta_{12}$, $\theta_{13}$ and $\Delta m^2_{21}$ fixed in both the true and test spectrum and varied the other parameters in the test spectrum. For values of $\delta_{\rm CP}$ in T2HK, we will consider the values of $0^\circ$ and $-90^\circ$. Note that as the mass ordering sensitivity of ICAL is only mildly dependent on $\delta_{\rm CP}$, and since simulating the ICAL data is  computationally expensive, we will only give results corresponding to $\delta_{\rm CP} = 0^\circ$ for ICAL. For $\theta_{23}$, we will consider the value as $45^\circ$ corresponding to maximal value of this mixing angle unless otherwise mentioned.  For JUNO, the oscillation probability is independent of $\theta_{23}$ and $\delta_{\rm CP}$. 

\section{Results}
\label{res}

In this section we will present our simulation results. We will first begin by discussing the main physics issue that is responsible for the synergy between the different experiments. In order to highlight it, we consider one experiment at a time and show how the best-fit value of $|\Delta m_{31}^2|$ comes out to be different for each oscillation channel. Since different oscillation channels are relevant for the different experiments, the $|\Delta m_{31}^2|$ that would give minimum $\chi^2$ is also different for them. As a result, when we combine data from two or three experiments and perform a joint analysis, the net mass ordering sensitivity obtained is significantly higher than what one would get by simply adding the individual $\chi^2$. 

\subsection{Mass Ordering Sensitivity - The Role of $|\Delta m^2_{31}|$}

As discussed before, in order to determine the statistical significance of the neutrino mass ordering in a given forthcoming experiment, we do the following. We simulate the prospective data for that experiment assuming, say, true normal ordering, and then fit it with a theory of inverted mass ordering. In the fit we allow the value of $|\Delta m^2_{31}|$ to vary freely within its current $3\sigma$ range. Hence, the minimum of the fit comes at a certain value of $|\Delta m^2_{31}|$ that is different from the one assumed in the data. In this subsection we highlight this aspect at the neutrino oscillation probability level. This is crucial to our understanding of the physics behind the synergy between the different experiments that we will see in the forthcoming sections. In particular, we will probe at what value of $|\Delta m^2_{31}|$ we get the minimum difference between the normal and inverted mass ordering in the oscillation probabilities for the different experiments. \\

\subsubsection{ICAL}

\begin{figure}[h!]
\centering
\parbox{7cm}{
\includegraphics[width=8cm]{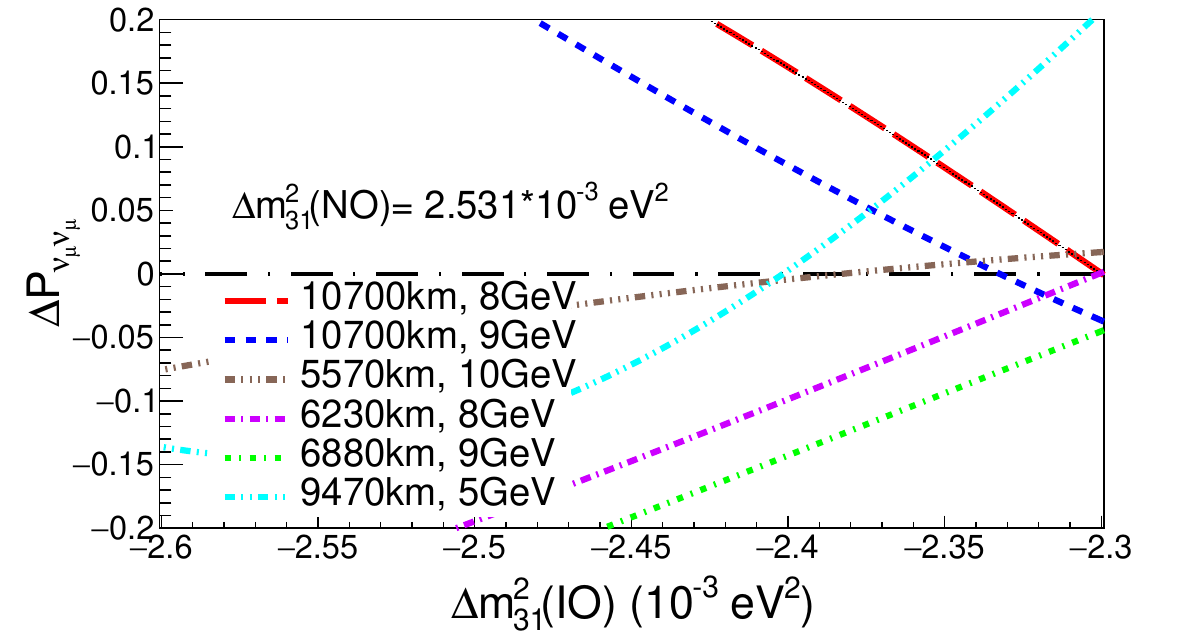}
 }
\qquad
\begin{minipage}{7cm}
\includegraphics[width=8cm]{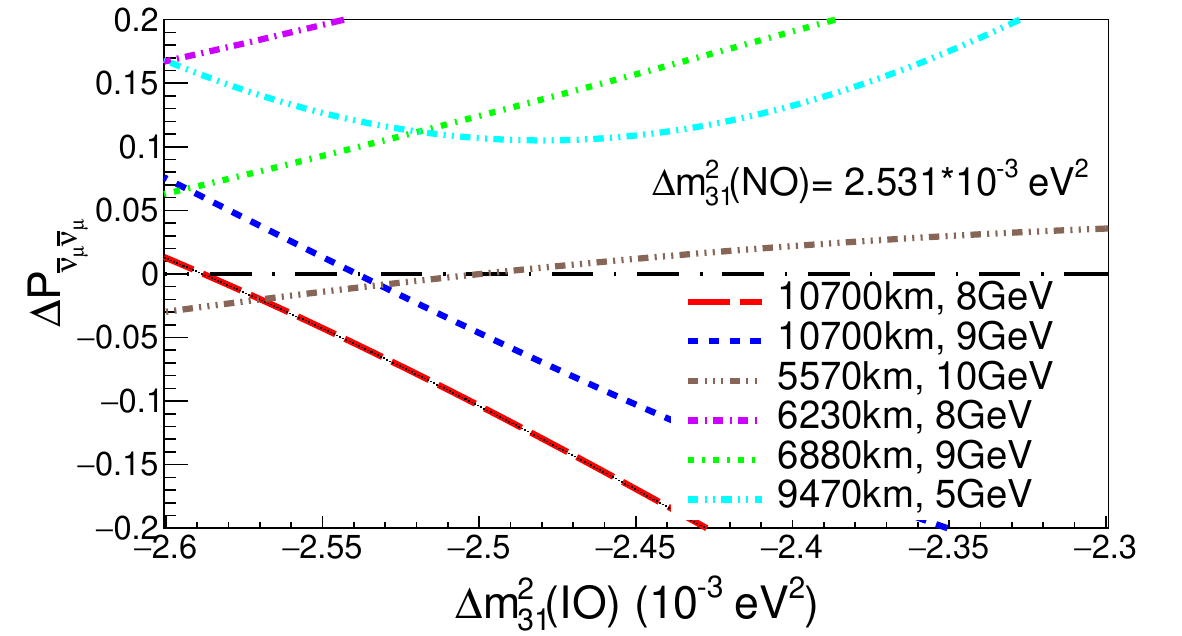}
\end{minipage}
\caption{Left panel shows the $P_{\nu_{\mu}\nu_{\mu}}(\rm NO)-P_{\nu_{\mu}\nu_{\mu}}$(IO) for neutrino and right panel shows the $ P_{\bar{\nu}_{\mu}\bar{\nu}_{\mu}}(\rm NO)-P_{\bar{\nu}_{\mu}\bar{\nu}_{\mu}}(\rm IO)$ for anti-neutrinos. These combinations of the energy and baseline have the maximum contribution towards the mass ordering sensitivity.}
\label{fig:inomin}
\end{figure}

\begin{figure}[h]
    \centering
    \includegraphics[width=0.6\textwidth]{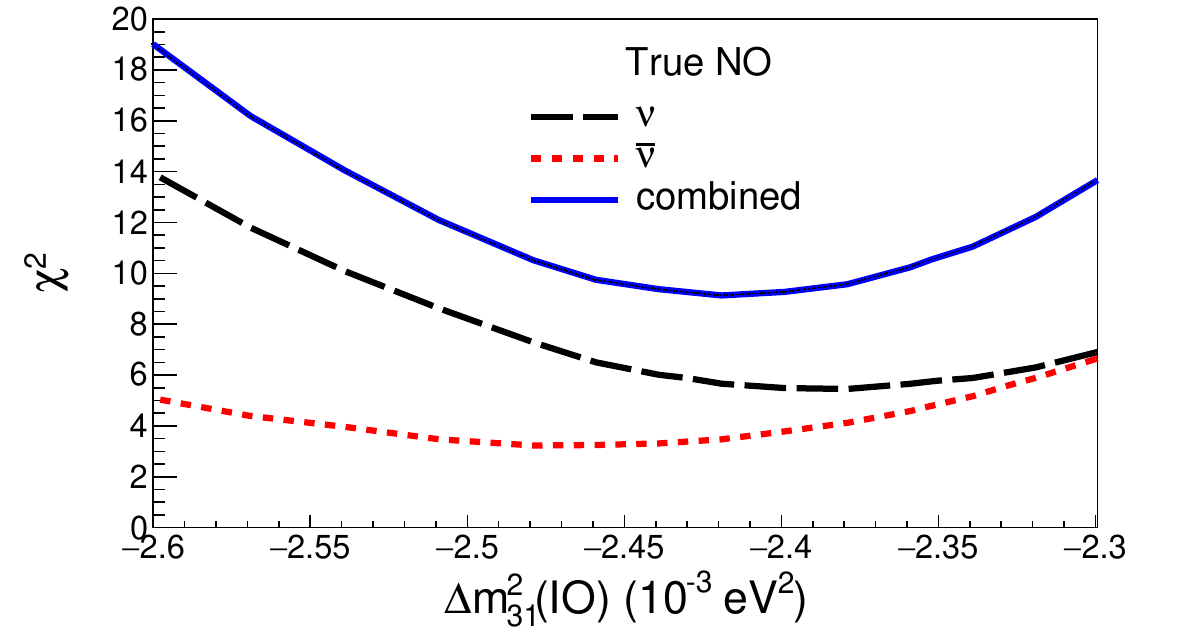}
    \caption{Mass ordering sensitivity $\chi^{2}$ as a function of  $\Delta m_{31}^2$(IO) for ICAL in $\nu$ , $\bar{\nu}$ and combined analysis. Red line is for $\bar{\nu}_{\mu}$ and black line for $\nu_{\mu}$ data and combined results are shown in blue lines.}
    \label{inochi}
\end{figure}

The proposed ICAL is an atmospheric neutrino experiment which will be sensitive to large energy range (0.5-25 GeV) and different baselines. In Fig.~\ref{fig:inomin}, we have plotted the difference between the disappearance channel probabilities for the two different mass orderings i.e. $\Delta P_{\mu \mu} = P_{\mu \mu} (\rm NO) - P_{\mu \mu} (\rm IO)$ in the $y$-axis. Note that in ICAL, the sensitivity to mass ordering mainly arises from the disappearance channel. In this figure, we have used the value of $\Delta m^{2}_{31} (\rm NO)=2.531 \times 10^{-3}$ eV$^2$ for normal ordering and the different values of $\Delta m^{2}_{31}$ (IO) for inverted ordering are plotted in the $x$-axis. The left panel is for neutrinos and the right panel is for antineutrinos. Different curves in the figure correspond to different combinations of baseline and energy, chosen because they have the maximum contribution towards the mass ordering sensitivity~\cite{Choubey:2005zy,Gandhi:2007td}. If the minimum value of  $\Delta P_{\mu \mu}$ occurs at different values of  $\Delta m^{2}_{31}$ (IO) for different bins, then it would give rise to synergy between these bins. From the figure we see that the minimum of $\Delta P_{\mu \mu}$ occurs at $\Delta m^{2}_{31} (\rm IO) \sim -2.3 \times 10^{-3}$ eV$^2$ (smaller values of $|\Delta m^{2}_{31}|$) for neutrinos and at $\Delta m^{2}_{31} (\rm IO) \sim -2.6 \times 10^{-3}$ eV$^2$ (larger values of $|\Delta m^{2}_{31}|$) for antineutrinos. This can also be seen in Fig.~\ref{inochi}, where we have plotted the mass ordering sensitivity $\chi^2$ for ICAL as a function of $\Delta m^{2}_{31}$(IO). The black curve corresponds to neutrinos and the red curve corresponds to antineutrinos. From these curves we see that for neutrinos the $\chi^2$ minimum is at $\Delta m^{2}_{31} (\rm IO) \simeq -2.380 \times 10^{-3}$ eV$^2$ and for the antineutrinos the minimum is at $\Delta m^{2}_{31} (\rm IO) \simeq -2.480 \times 10^{-3}$ eV$^2$. This is in agreement with what we concluded from the probability level discussions. The combined sensitivity coming from the neutrino data and antineutrino data is shown by the blue curve in Fig.~\ref{inochi}. From this curve we note that when the neutrino and antineutrino sensitivities are combined, the $\chi^2$ minimum appears around $\Delta m^{2}_{31} (\rm IO) = -2.419 \times 10^{-3}$ eV$^2$, which is different from the true value of  $\Delta m^{2}_{31} (\rm NO) = 2.531 \times 10^{-3}$ eV$^2$.

\subsubsection{T2HK and JUNO} 

The dependence of $\chi^2$ minimum with respect to $\Delta m^{2}_{31}$ (IO) for T2HK and JUNO was recently studied in detail by us in Ref.~\cite{Choubey:2022gzv}. We explored the synergy 
\begin{itemize}
    \item between the appearance and the disappearance channels of T2HK,
    \item between the different $E$ bins of JUNO, and
    \item between T2HK and JUNO.
\end{itemize}
We summarise here the main points for completeness. T2HK has a baseline of 295 km and a narrow-band beam peaked at neutrino energy of 0.6 GeV. Therefore, (1) low $E$ ensures that matter effects in the disappearance channel of T2HK can be neglected for all practical purposes and (2) the narrow-band beam ensures that we can essentially work at the oscillation maxima. It can be shown that for a true value of $\Delta m^{2}_{31}$(NO), the $\chi^2$ minimum for $\Delta m^{2}_{31}$(IO) should appear at
\begin{equation}
\Delta m_{31}^{2}({\rm IO})=-\Delta m_{31}^{2}({\rm NO}) + x ,
\end{equation}
with
\begin{equation}
x = \frac{2|U_{\mu 2}|^{2}}{|U_{\mu 1}|^{2}+|U_{\mu 2}|^{2}}\Delta m_{21}^{2},  
\end{equation}
where $U$ is the PMNS matrix. For our choice of oscillation parameters, we obtain 
\begin{eqnarray}
\Delta m_{31}^{2}(\rm IO) = -2.45\times10^{-3} ~{\rm {eV}}^{2}. 
\label{eq:bfapprox}
\end{eqnarray}
It was shown in~\cite{Choubey:2022gzv} that both at the probability level as well as the $\chi^2$ level, the minima for the disappearance channel comes very close to the value obtained in Eq.~(\ref{eq:bfapprox}). Indeed the mass ordering $\chi^2$ goes to zero at the above value of $\Delta m^{2}_{31}$(IO). The appearance channel on the other hand, has matter effects and hence for this channel the mass ordering sensitivity never goes to zero. Nevertheless it too has a distinct minima with respect to $\Delta m^{2}_{31}$(IO), which is very different from the one for the disappearance channel. Hence, when we combine the appearance and the disappearance channels, then the synergy coming from different values of $\Delta m^{2}_{31}$(IO) at which we get the individual $\chi^2$ minima results in a significantly enhanced $\chi^2$ for T2HK. 

In JUNO the situation is more complex due to its wide energy range of 1.8 MeV to 8.0 MeV. The condition for minima is given by the relation \cite{Choubey:2022gzv} 
\begin{eqnarray}
&& \cos^2\theta_{12}\sin 2\Delta_{31}^{IO}+\sin^2\theta_{12}\sin2(\Delta_{31}^{IO}-\Delta_{21}) \nonumber \\ \nonumber
&-&\cos^2\theta_{12}\sin 2\Delta _{31}^{NO}.\frac{\Delta_{31}^{NO}}{\Delta_{31}^{IO}}-\sin^2\theta_{12}\sin2\Delta_{32}^{NO}\frac{\Delta_{32}^{NO}}{\Delta_{31}^{IO}}=0    ,
\label{eq:minrelation}
\end{eqnarray}
where $\Delta_{ij}=\Delta m_{ij}^{2}L/4E$. Hence, every energy bin of JUNO has a mass ordering minimum of $\simeq 0$ but at a different value of $\Delta m^{2}_{31}$(IO). This leads to a synergy between the 200 bins and when we do the full spectral analysis, we get a sizable sensitivity to mass ordering. 

From the above discussion we see that all the three experiments ICAL, T2HK and JUNO, exhibit different best-fit values of $\Delta m^{2}_{31}$(IO)  for  the same true value of $\Delta m^{2}_{31}$(NO). For this reason, when these experiments are combined, we expect a significant improvement in the overall mass ordering sensitivity.

\subsection{Combined sensitivity of ICAL and JUNO}

\begin{figure}[h]
    \centering
    \includegraphics[width=0.6\textwidth]{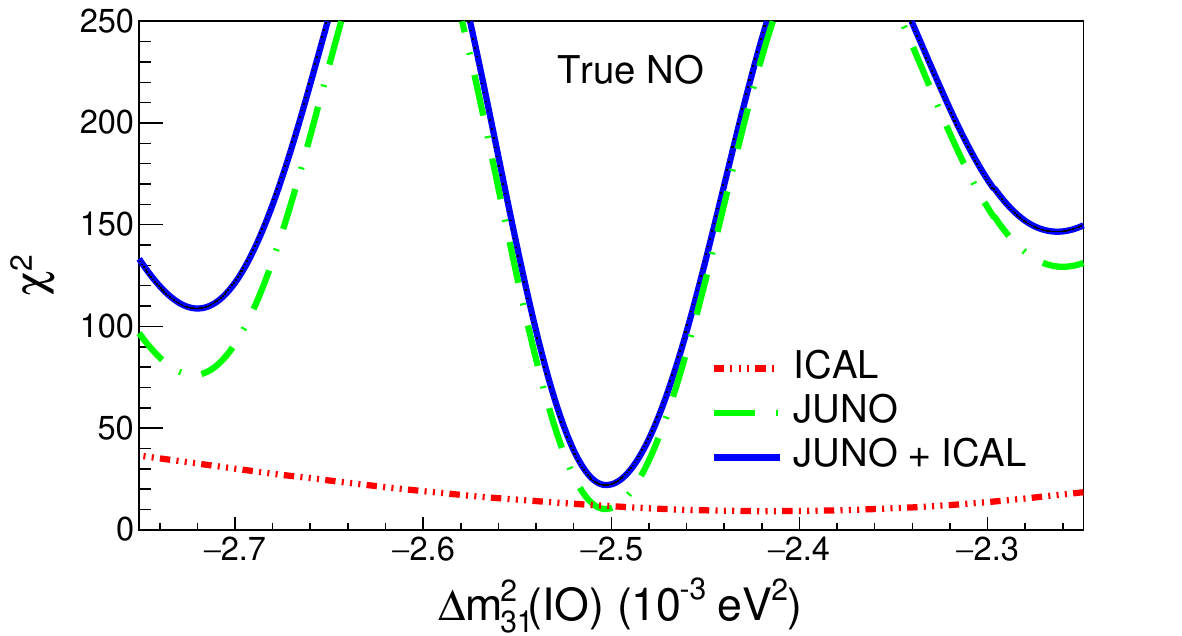}
    \caption{Mass ordering sensitivity expected from a combined analysis of JUNO and ICAL, shown as a function of $\Delta m^2_{31}$(IO).}
    \label{fig:inojuno}
\end{figure}

We start with the combined sensitivity of ICAL and JUNO shown in Fig.~\ref{fig:inojuno}. In this figure (and all the subsequent figures) the lines corresponding to the different experiments and different combination of experiments are clearly marked on the figure. The analysis procedure used was outlined in the previous section. We have plotted the y-axis up to $\chi^2 = 250$ to show the multiple local minima of JUNO. The sensitivity of ICAL alone for 10 years exposure is shown as a function of $\Delta m_{31}^2$(IO) by the red line. We notice that the dependence of the ICAL mass-ordering $\chi^2$ on $\Delta m_{31}^2$(IO) is significantly shallow in contrast to the one of JUNO, shown by the green line. The mass-ordering $\chi^2$ obtained from the combined analysis of ICAL and JUNO is shown by the blue line. We note that the best-fit $\Delta m_{31}^2$(IO) of the combined  fit is mostly determined by JUNO, owing to its sharp dependence on  $\Delta m_{31}^2$(IO). However, the inclusion of ICAL does play a role and brings a synergy, which raises the overall $\chi^2$. The exact values of $\Delta m_{31}^2$(IO) for which the mass-ordering $\chi^2$ is minimum is given in Table~\ref{tab:inojunobf} while the $\chi^2$ minimum values are given in Table~\ref{tab:inojunochi}. We see that the combined analysis gives us about 14\% increase in the sensitivity as compared to a simple sum of the $\chi^2$ of the two experiments. 

\begin{table}[h!]
\centering
\begin{tabular}{|| c | c | c |c ||}
\hline
$\Delta m^{2}_{31}$ (True) & $\Delta m^{2}_{31}$ (JUNO) & $\Delta m^{2}_{31}$ (ICAL)   & $\Delta m^{2}_{31}$ (Combined)   \\
\hline 
  2.531&-2.503 & -2.419 & -2.503 \\
\hline
\end{tabular}
\caption{Column 1 shows the true value of $\Delta m_{31}^2$(NO) (in units of $10^{-3}$ eV$^{2}$) while columns 2, 3 and 4 show values of $\Delta m_{31}^{2}$(IO) (in units of $10^{-3}$ eV$^{2}$), for which we get mass-ordering $\chi^2$ minimum for ICAL, JUNO and (ICAL and T2HK) Combined.}
\label{tab:inojunobf}
\end{table}
\begin{table}[h!]
\centering
\begin{tabular}{|| c | c | c |c | c ||}
\hline
$\chisq_{\rm JUNO}$ & $\chisq_{\rm ICAL}$ & $\chisq_{\rm JUNO}+\chisq_{\rm ICAL}$ & $\chisq_{\rm Combined}$ & $\%$ increase\\
\hline  
 10.23 & 9.12 & 19.35 & 22.01 & 13.7\\
\hline
\end{tabular}
\caption{Mass-ordering $\chi^2$ for true NO and test IO. The last column shows the percentage increase in sensitivity we get from a combined analysis (column 4) as compared to a simple sum (column 3).  
}
\label{tab:inojunochi}
\end{table}

\subsection{Combined sensitivity of ICAL and T2HK}

\begin{figure}[h]
\centering
\parbox{7cm}{
\includegraphics[width=8cm]{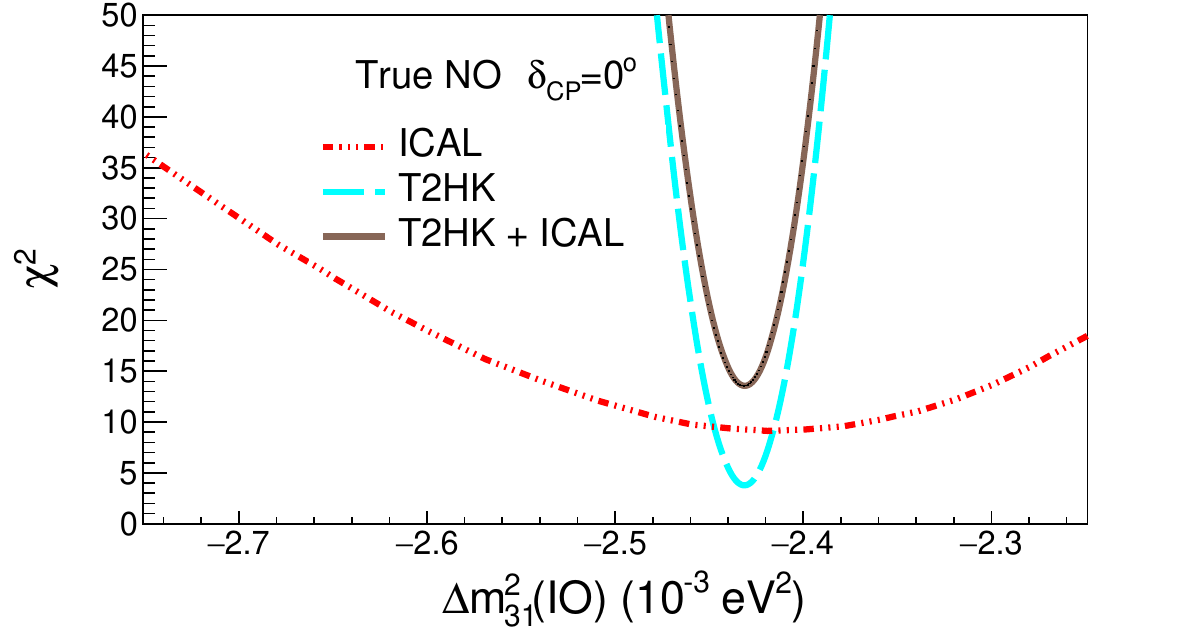}
 }
\qquad
\begin{minipage}{7cm}
\includegraphics[width=8cm]{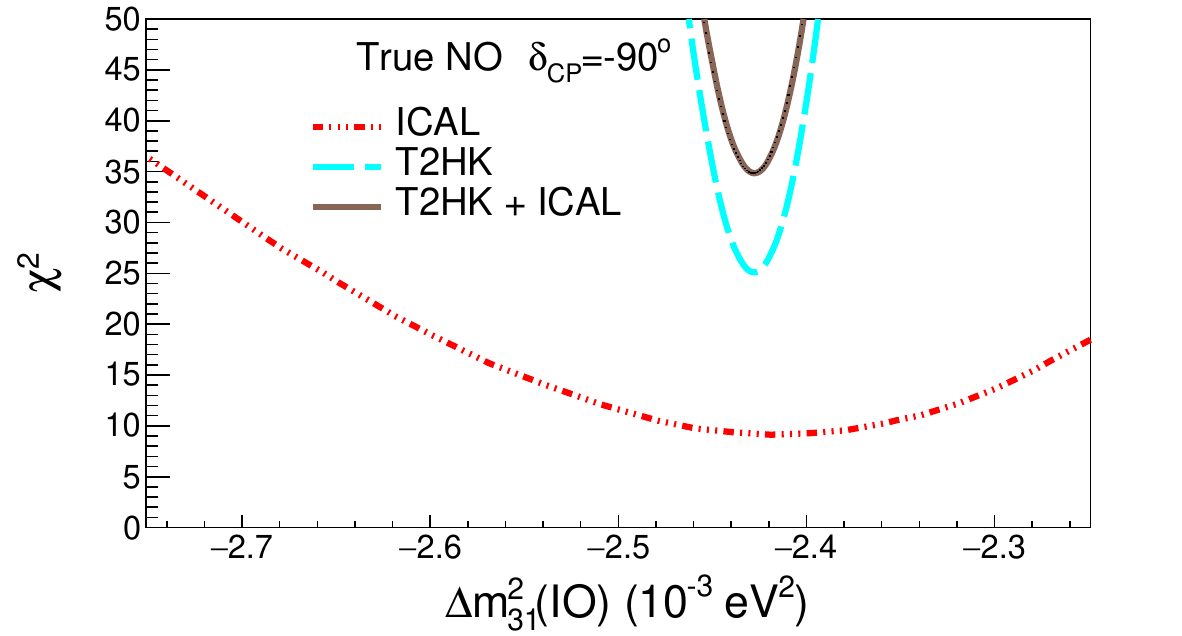}
\end{minipage}
\caption{Mass ordering sensitivity as a function of $\Delta m^2_{31}$(IO). Left panel is for $\delta_{\rm CP} = 0^\circ$ and the right panel is for $\delta_{\rm CP} =-90^\circ$ corresponding to T2HK.}
\label{t2hkino}
\end{figure}

Next we perform a combined study of long baseline experiment T2HK and atmospheric experiment ICAL. The results are shown in Fig.~\ref{t2hkino}. As seen before, ICAL gives us 3$\sigma$ mass ordering sensitivity with 10 year data for $\theta_{23}$ maximal. T2HK also gives us mass ordering sensitivity but it is $\delta_{\rm CP}$ dependent (Fig.~\ref{t2hkino}). This leads to the well known hierarchy - $\delta_{\rm CP}$ degeneracy \cite{Barger:2001yr,Prakash:2012az,Ghosh:2015ena}. In T2HK we get mass ordering sensitivity below 2$\sigma$ for $\delta_{\rm CP}=0^{\circ}$ (left panel) and 5$\sigma$ for $\delta_{\rm CP}=-90^{\circ}$ (right panel). When we combine ICAL and T2HK data we get 3.68$\sigma$ sensitivity for $\delta_{\rm CP}=0^{\circ}$ and 5.9$\sigma$ for $\delta_{\rm CP}=-90^{\circ}$ . Synergy effect between ICAL and T2HK is very small. The reason is that the combined best-fit value of $\Delta m^{2}_{31}(\rm IO)$ is governed almost entirely by the T2HK and INO is unable to make much impact on it. Thus the synergistic effect is very weak. 
\begin{table}[h!]
\centering
\scalebox{0.8}{
\begin{tabular}{||c| c  | c | c |c ||}
\hline
& $\Delta m^{2}_{31}$ (True) & $\Delta m^{2}_{31}$ (ICAL)  & $\Delta m^{2}_{31}$ (T2HK)   & $\Delta m^{2}_{31}$ (Combined)   \\
\hline 
$\delta_{\rm CP}=0$ &2.531 & -2.419  & -2.431 & -2.431  \\
\hline
$\delta_{\rm CP}=-90$ & 2.531 & -2.419  & -2.428 & -2.428 \\
\hline
\end{tabular}}
\caption{Column 2 shows the true value of $\Delta m_{31}^2$(NO) (in units of $10^{-3}$ eV$^{2}$) while columns 3, 4 and 5 show values of $\Delta m_{31}^{2}$(IO) (in units of $10^{-3}$ eV$^{2}$), for which we get mass-ordering $\chi^2$ minimum for ICAL, T2HK and (ICAL and T2HK) Combined.}
\label{tab:inot2hkbf}
\end{table}
\begin{table}[h!]
\centering
\begin{tabular}{||c|c | c | c |c | c |c ||}
\hline
& $\chisq_{\rm ICAL}$  & $\chisq_{\rm T2HK}$ &$\chisq_{\rm T2HK}+\chisq_{\rm ICAL}$ & $\chisq_{\rm Combined}$ & $\%$ increase\\
\hline 
$\delta_{\rm CP}=0$ & 9.12   & 3.77 & 12.89 &13.55 & 5 \\
\hline
$\delta_{\rm CP}=-90$ & 9.12   & 25.09 & 34.21 & 34.84 & 1.8\\
\hline
\end{tabular}
\caption{Mass-ordering $\chi^2$ for true NO and test IO. The last column shows the percentage increase in sensitivity we get from a combined analysis (column 5) as compared to a simple sum (column 4). }
\label{tab:inot2hkchi}
\end{table}
We can see in Fig.~\ref{t2hkino} and Table~\ref{tab:inot2hkbf} that the best-fit values are similar for both experiments. Also, ICAL has a very flat curve corresponding to $\Delta m^{2}_{31}$(IO). So this leads to very small synergy effect as can be seen in Table \ref{tab:inot2hkchi}. The percentage increase as compared to a simple sum of the $\chi^{2}$ is 5$\%$ for $\delta_{\rm CP}=0^{\circ}$ and 2$\%$ for $\delta_{\rm CP}=-90^{\circ}$. However, it is important to note that by combining T2HK and ICAL, the sensitivity to mass ordering improves to about $4 \sigma$ even for $\delta_{\rm CP} = 0^\circ$, while for $\delta_{\rm CP}=-90^{\circ}$ we have close to $6\sigma$ sensitivity.

 \subsection{Combined sensitivity of ICAL, JUNO and T2HK}

\begin{figure}[h]
\centering
\parbox{7cm}{
\includegraphics[width=8cm]{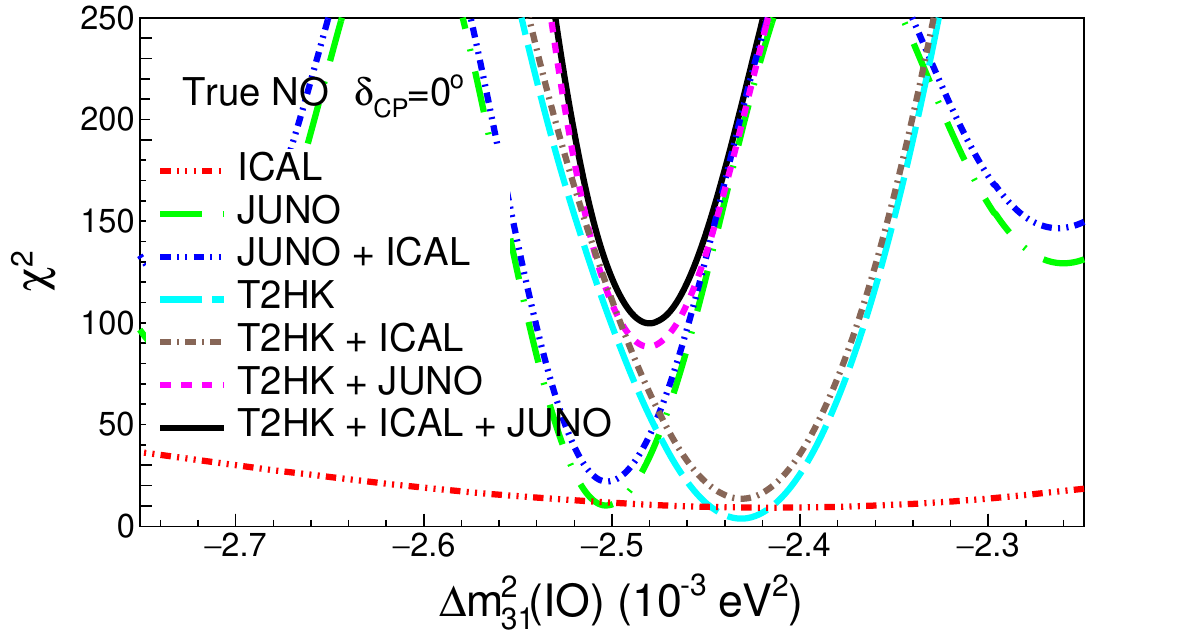}
 }
\qquad
\begin{minipage}{7cm}
\includegraphics[width=8cm]{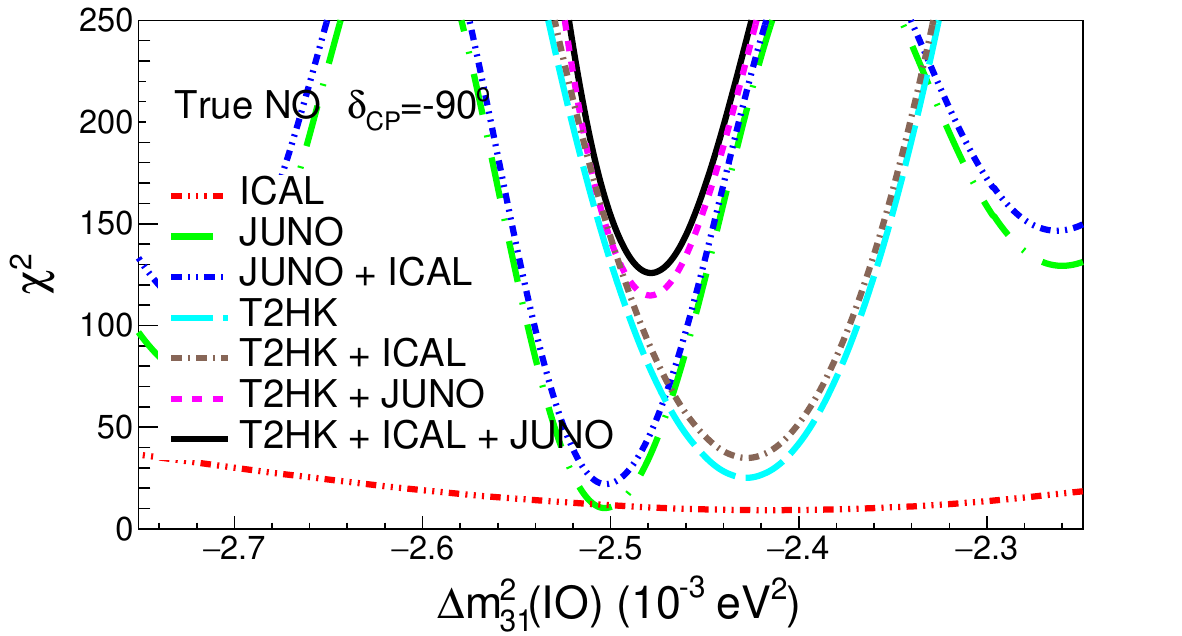}
\end{minipage}
\caption{Mass ordering sensitivity as a function of $\Delta m^2_{31}$(IO). Left panel is for $\delta_{\rm CP} = 0^\circ$ and the right panel is for $\delta_{\rm CP} =-90^\circ$ corresponding to T2HK.}
\label{combined}
\end{figure}

In this section we will see that by combining atmospheric data (ICAL) with long baseline (T2HK) and reactor (JUNO) data, we can achieve more than 10 $\sigma$ sensitivity for $\delta_{\rm CP}=0^\circ$ and 11 $\sigma$ sensitivity for $\delta_{\rm CP}=-90^\circ$. The results are presented in Fig. \ref{combined}, as a function of $\Delta m^2_{31}$ (IO). In the left (right) panel we have considered $\delta_{\rm CP} = 0^\circ ~ (-90^\circ)$. The values of $\Delta m_{31}^2$(IO) for which the mass-ordering $\chi^2$ is minimum is given in Table~\ref{tab:inojunot2hkbf} while the $\chi^2$ minimum values are given in Table~\ref{tab:inojunot2hkchi}.

\begin{table}[h!]
\centering
\scalebox{0.8}{
\begin{tabular}{||c| c |c | c |c|c ||}
\hline
& $\Delta m^{2}_{31}$ (True) & $\Delta m^{2}_{31}$ (ICAL) & $\Delta m^{2}_{31}$ (JUNO) & $\Delta m^{2}_{31}$ (T2HK)   & $\Delta m^{2}_{31}$ (Combined)  \\
\hline 
$\delta_{\rm CP}=0^\circ$ & 2.531& -2.419 & -2.503 & -2.431 & -2.48 \\
\hline
$\delta_{\rm CP}=-90^\circ$ & 2.531& -2.419 & -2.503 & -2.428 & -2.479 \\
\hline
\end{tabular}}
\caption{Column 2 shows the true value of $\Delta m_{31}^2$(NO) (in units of $10^{-3}$ eV$^{2}$) while columns 3, 4, 5 and 6 show values of $\Delta m_{31}^{2}$(IO) (in units of $10^{-3}$ eV$^{2}$), for which we get mass-ordering $\chi^2$ minimum for ICAL, JUNO, T2HK and (ICAL and JUNO and T2HK) Combined.
}
\label{tab:inojunot2hkbf}
\end{table}

\begin{table}[h!]
\centering
\scalebox{0.8}{
\begin{tabular}{||c| c |c | c | c |c | c |c ||}
\hline
& $\chisq_{\rm ICAL}$ & $\chisq_{\rm JUNO}$ & $\chisq_{\rm T2HK}$ & $\chisq_{\rm JUNO}+\chisq_{\rm T2HK}+\chisq_{\rm ICAL}$ & $\chisq_{\rm Combined}$ & $\%$ increase\\
\hline 
$\delta_{\rm CP}=0^\circ$ & 9.12 & 10.23 & 3.77 & 23.12 &99.76 & 331 \\
\hline
$\delta_{\rm CP}=-90^\circ$ & 9.12  & 10.23 & 25.09 & 44.44 & 125.79 & 183\\
\hline
\end{tabular}}
\caption{Mass-ordering $\chi^2$ for true NO and test IO. The last column shows the percentage increase in sensitivity we get from a combined analysis (column 6) as compared to a simple sum (column 5). }
\label{tab:inojunot2hkchi}
\end{table}
 
As we discussed earlier, for JUNO the mass-ordering $\chi^2$ is about 10 and for ICAL the mass ordering sensitivity is about 9. The sensitivity of T2HK is about 4 for $\delta_{\rm CP} = 0^\circ$ and 25 for $\delta_{\rm CP} = -90^\circ$. We again stress upon the fact that though the true value of $\Delta m^2_{31}$ is same for all these experiments, the $\chi^2$ minimum occurs at different values of $\Delta m^2_{31}$ (IO) for different experiments.  
 For the combination of ICAL, JUNO and T2HK, the mass-ordering $\chi^2$ increases to about 100 (126) for $\delta_{\rm CP} = 0^\circ (-90^\circ)$. 
 We had seen that for the combination of JUNO+ICAL and T2HK+ICAL, the improvement due to synergy was rather mild. We had explained that this was because the ICAL $\chi^2$ was shallow and was therefore unable to really change the best-fit $\Delta m_{31}^2$(IO) significantly. On the other hand, both JUNO and T2HK have well-defined and sharp $\chi^2$ with best-fit  $\Delta m_{31}^2$(IO) distinctly different.  This was discussed in Ref.~\cite{Choubey:2022gzv}. As a result, when we combined data from all experiments, we get a big synergistic boost leading to the staggering improvement as can be seen in the last column of Table~\ref{tab:inojunot2hkchi}.

\subsection{Effect of energy resolution of JUNO}

\begin{figure}[h]
    \centering
    \includegraphics[width=0.6\textwidth]{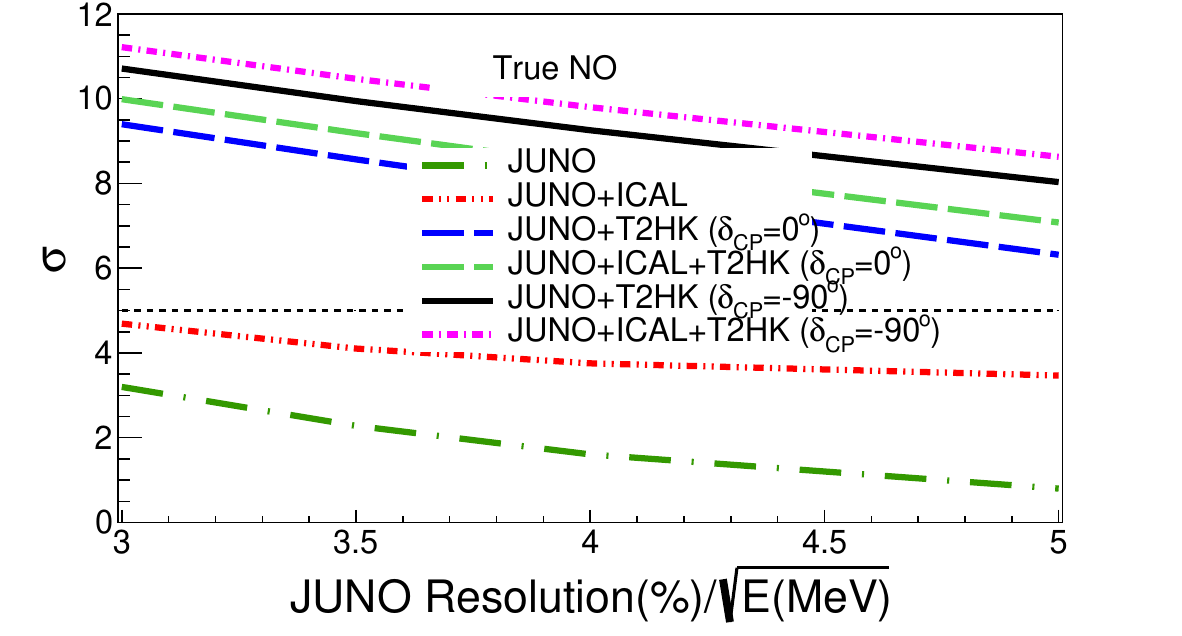}
    \caption{Neutrino mass ordering sensitivity as a function of energy resolution of JUNO.}
    \label{fig:mesh1}
\end{figure}
Next we discuss how the mass ordering sensitivity changes if we vary the energy resolution of JUNO. In Fig. \ref{fig:mesh1}, we have presented the mass-ordering sensitivity as a function of the energy resolution of JUNO. From this figure we clearly see that the sensitivity of JUNO goes from $\chi^2 = 10$ to less than 1 when the energy resolution is varied from 3$\%/\sqrt{E(MeV)}$ to 5$\%/\sqrt{E(MeV)}$. But when it is combined with other experiments we can see that the sensitivity is significantly improved. For ICAL+JUNO+T2HK the $\chi^2$ is expected to come around 52 (76) for $\delta_{\rm CP} = 0^\circ (-90^\circ)$ even if the JUNO energy resolution is 5$\%/\sqrt{E(MeV)}$. This shows that even for poorer resolutions of JUNO, with the combination of T2HK and ICAL, one can achieve a mass ordering sensitivity of $7.2 \sigma$ even for the unfavourable value of $\delta_{\rm CP}$. For JUNO+ICAL, the sensitivity is expected to be around $4 \sigma$ for JUNO energy resolution of 5$\%/\sqrt{E(MeV)}$.

\subsection{Effect of ICAL run-time}

\begin{figure}[h]
    \centering
    \includegraphics[width=0.6\textwidth]{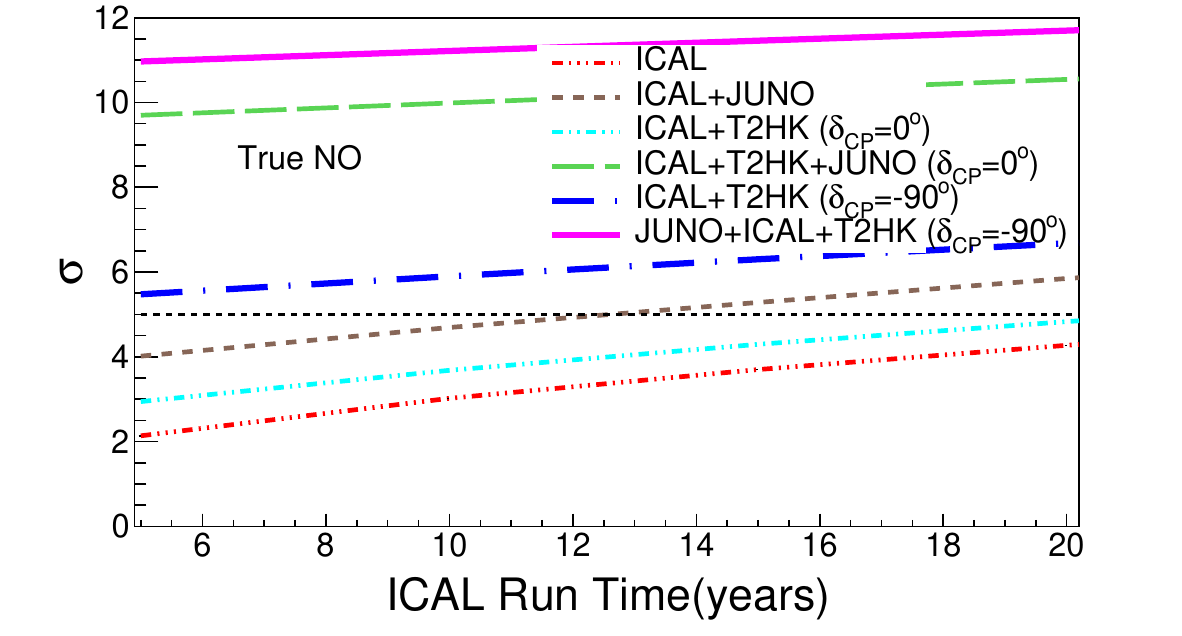}
    \caption{Neutrino mass ordering sensitivity as a function of run-time of ICAL.}
    \label{fig:mesh2}
\end{figure}
In this section we explore the effect of extending the run-time of ICAL on the mass ordering sensitivity. As atmospheric neutrinos are available without any cost \footnote{For the detector of course there will be running costs such as maintaining the magnetic field, electronics and power supply for cooling.}, one can in principle run an atmospheric neutrino experiment for a longer time. In Fig. \ref{fig:mesh2}, we present the mass-ordering sensitivity as a function of run-time of ICAL. For a 20 years running of ICAL, the $\chi^2$ for ICAL goes up to 16. When it is added with JUNO the $\chi^2$ reaches 36. We can see that $5\sigma$ sensitivity for mass ordering could be achieved with just JUNO and 12 years of running of ICAL. For the combination of ICAL+JUNO+T2HK one can have a mass-ordering $\chi^2$ of 100 (144) for $\delta_{\rm CP} = 0^\circ (-90^\circ)$ for 20 years running of ICAL. 

\subsection{Effect of true values $\theta_{23}$ on combined analysis}
 
\begin{figure}[h!]
\centering
\parbox{7cm}{
\includegraphics[width=8cm]{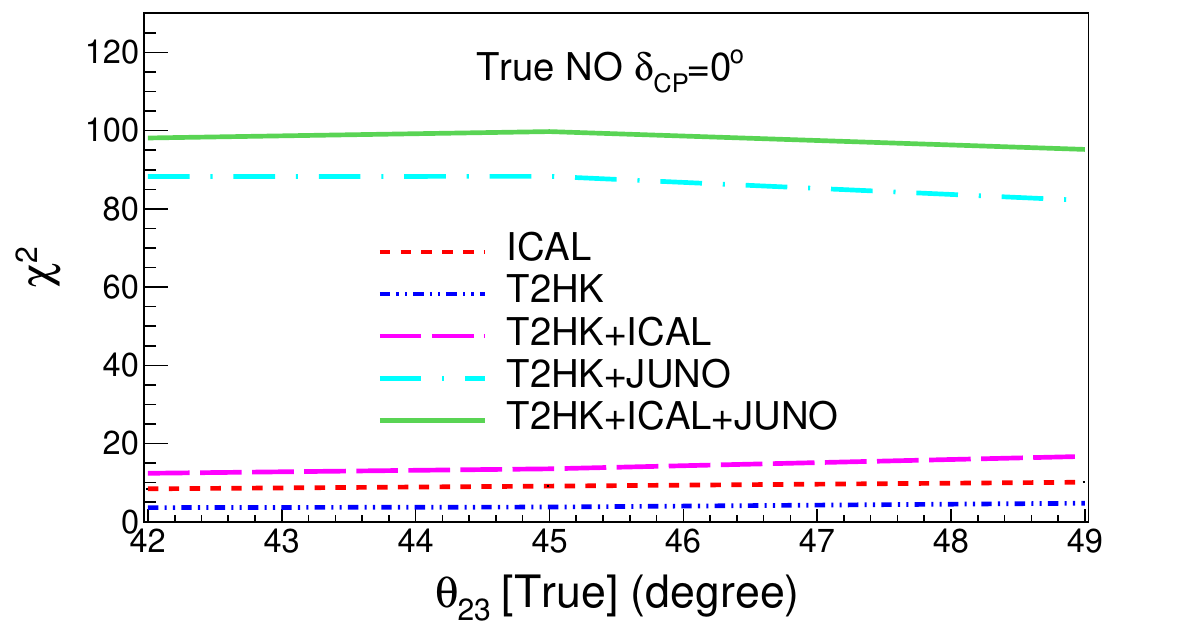}
 }
\qquad
\begin{minipage}{7cm}
\includegraphics[width=8cm]{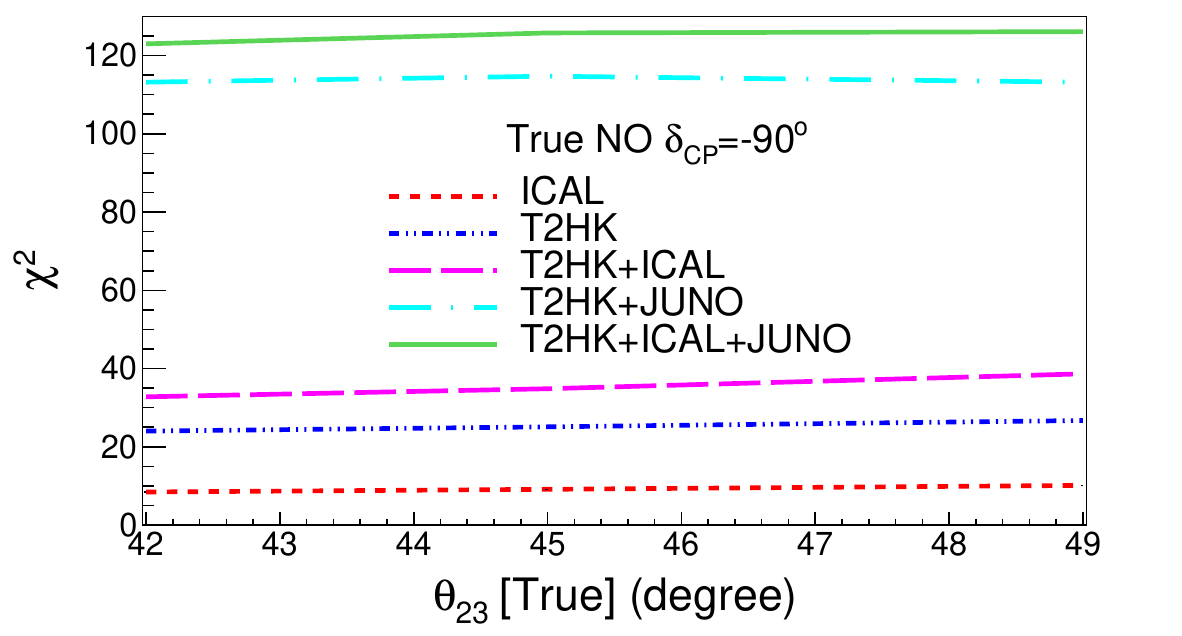}
\end{minipage}
\caption{Neutrino mass ordering sensitivity as a function of $\theta_{23}$ (true). Left panel is for $\delta_{\rm CP} = 0^\circ$ and the right panel is for $\delta_{\rm CP} =-90^\circ$ corresponding to T2HK.}
\label{fig:2figsB_1}
\end{figure}
\begin{figure}[h!]
\centering
\parbox{7cm}{
\includegraphics[width=8cm]{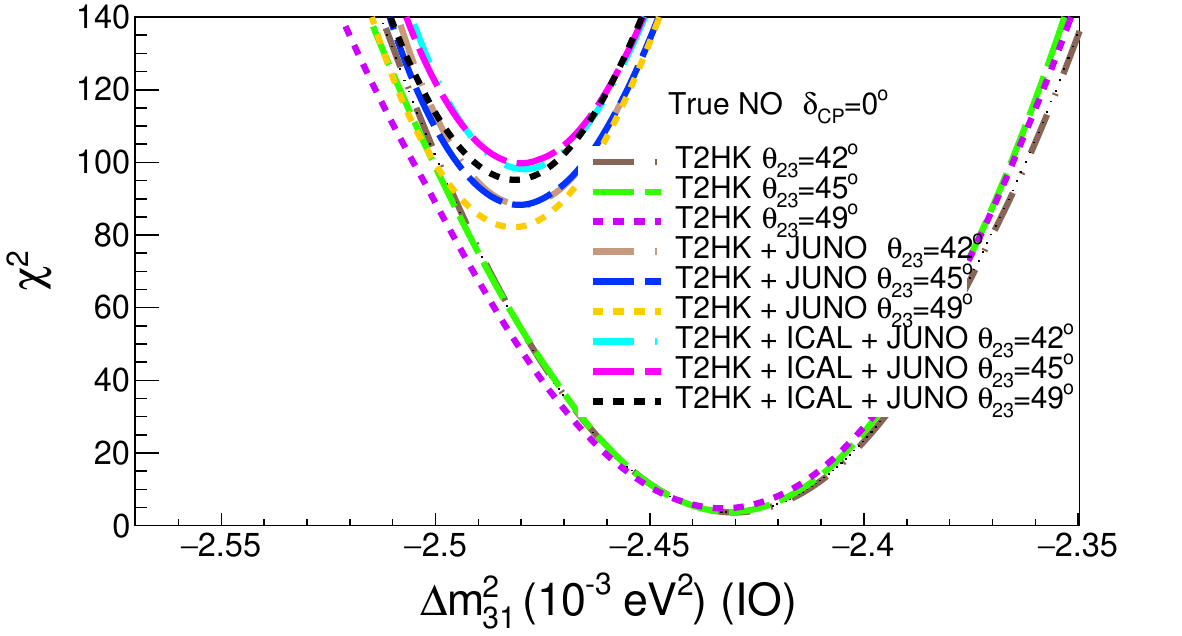}
 }
\qquad
\begin{minipage}{7cm}
\includegraphics[width=8cm]{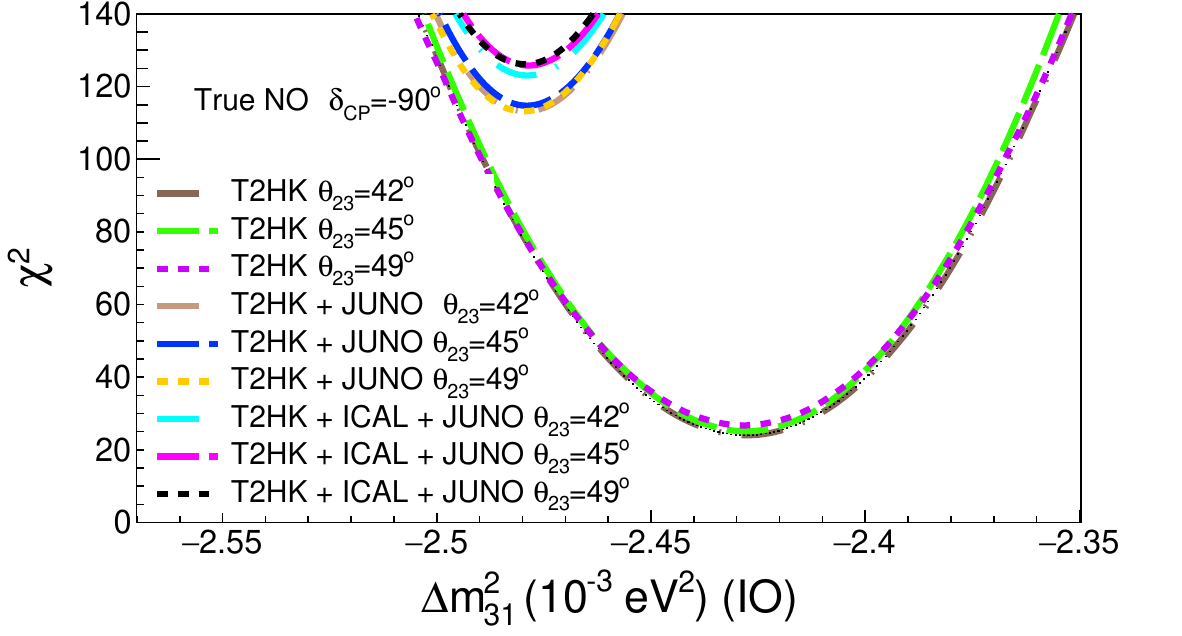}

\end{minipage}
\caption{Mass ordering sensitivity as a function of $\Delta m^2_{31}$(IO) for different values of $\theta_{23}$. Left panel is for $\delta_{\rm CP} = 0^\circ$ and the right panel is for $\delta_{\rm CP} =-90^\circ$ corresponding to T2HK.}
\label{fig:2figsB_2}
\end{figure}

Now let us see how the mass ordering sensitivity varies when we vary the true value of $\theta_{23}$. In Fig. \ref{fig:2figsB_1} we have plotted the mass-ordering $\chi^2$ as a function of $\theta_{23}$ (true). The value of $\delta_{\rm CP}$ for T2HK is $0^\circ$ in the left panel and $-90^\circ$ in the right panel. In general as $\theta_{23}$ increases, the mass ordering sensitivity increases ~\cite{anu-mh, moon:2014}. We can see this general trend in most of the curves, except for the T2HK+JUNO curve for both the values of $\delta_{\rm CP}$ and ICAL+JUNO+T2HK curve for $\delta_{\rm CP} = 0^\circ$. For these curves, the $\chi^2$ increases as $\theta_{23}$ increases from $42^\circ$ to $45^\circ$ and when $\theta_{23}$ increases further from $45^\circ$, the sensitivity decreases. To understand this, in Fig. \ref{fig:2figsB_2}, we have plotted $\chi^2$ vs $\Delta m^2_{31}$(IO) for three true values of $\theta_{23}$ i.e., $42^\circ$, $45^\circ$ and $49^\circ$.
The value of $\delta_{\rm CP}$ for T2HK is $0^\circ$ in the left panel and $-90^\circ$ in the right panel.
\begin{itemize}
    \item For $\delta_{\rm CP} = 0^\circ$, we note that though $\chi^2$ for T2HK increases as $\theta_{23}$ increases, the $\chi^2$ minimum tends to shift towards the left. As a result when it is combined with JUNO, the $\chi^2$ corresponding to $\theta_{23} = 49^\circ$ becomes lower than the $\chi^2$ for $\theta_{23}=42^\circ$. However for ICAL+JUNO+T2HK, the $\chi^2$ for $\theta_{23} = 49^\circ$ is higher than $\chi^2$ for $\theta_{23} = 42^\circ$ but lower than $\chi^2$ for $\theta_{23} = 45^\circ$. 
    
    \item For $\delta_{\rm CP} = -90^\circ$, the shift of the T2HK curve towards left for the value of $\theta_{23} = 49^\circ$ is less prominent as compared to $\delta_{\rm CP} = 0^\circ$. As a result, for T2HK+JUNO, the $\chi^2$ for $\theta_{23} = 49^\circ$ is higher than $\chi^2$ for $\theta_{23} = 42^\circ$ but lower than $\chi^2$ for $\theta_{23} = 45^\circ$. However, for the combination of ICAL+JUNO+T2HK, the $\chi^2$ for $\theta_{23} = 49^\circ$ is higher than $\chi^2$ for $\theta_{23} = 42^\circ$ and $45^\circ$ establishing the general trend.
\end{itemize}

\subsection{Effect of octant degeneracy in ICAL and T2HK on mass ordering }
 
\begin{figure}[h!]
\centering
\parbox{7cm}{
\includegraphics[width=8cm]{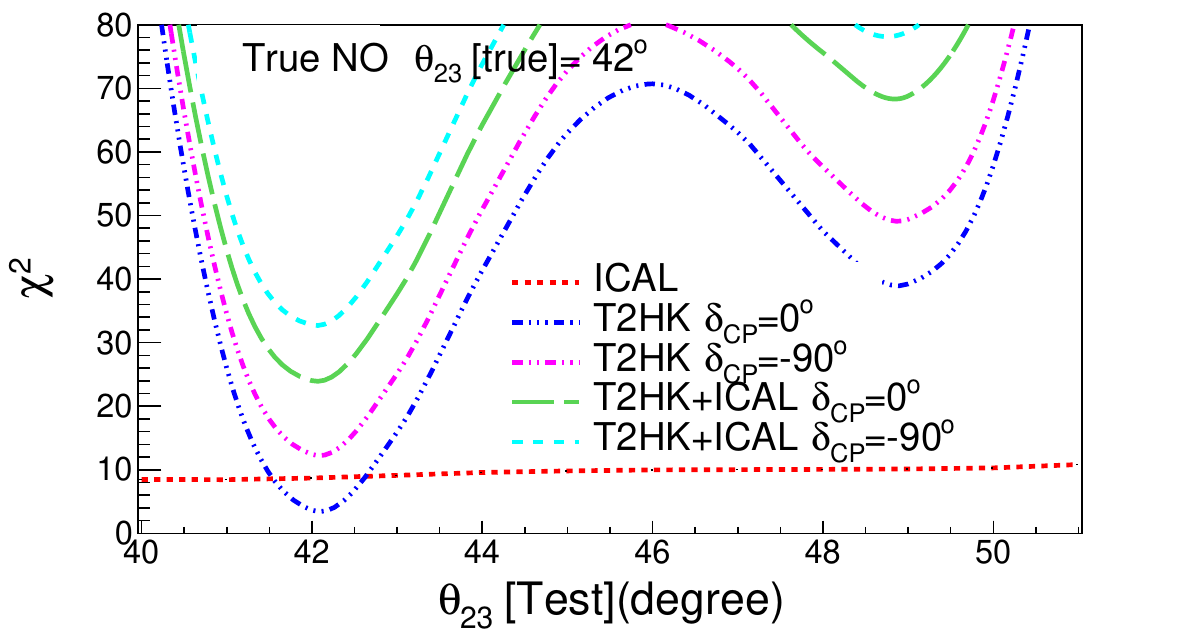}
 }
\qquad
\begin{minipage}{7cm}
\includegraphics[width=8cm]{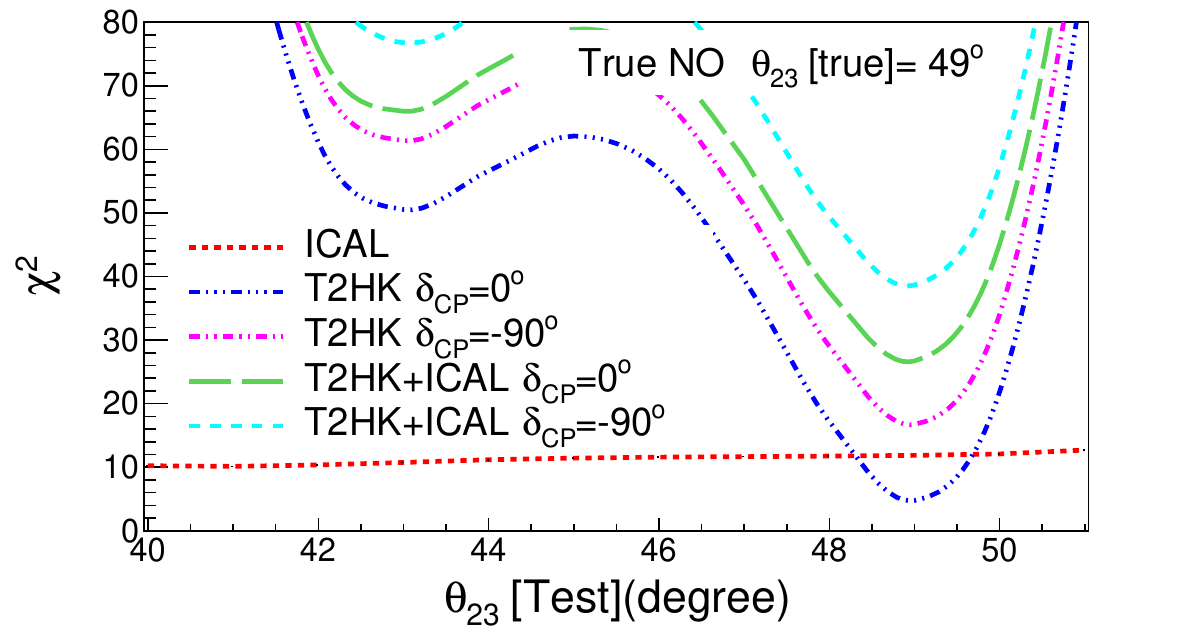}

\end{minipage}
\caption{Neutrino mass ordering sensitivity as a function of $\theta_{23}$ (test). Left panel is for $\delta_{\rm CP} = 0^\circ$ and the right panel is for $\delta_{\rm CP} =-90^\circ$ corresponding to T2HK.}
\label{fig:2figsB_3}
\end{figure}
Finally let us discuss the effect of octant degeneracy in the mass ordering sensitivity of T2HK and ICAL. In Fig. \ref{fig:2figsB_3}, we have plotted the mass-ordering $\chi^2$ as a function of $\theta_{23}$ (test). The left panel is for $\theta_{23}$ (true) $= 42^\circ$ and the right panel is for $\theta_{23}$ (true) $= 49^\circ$. From the panels we see that for true $\theta_{23}$ in the lower octant, the $\chi^2$ minimum always occurs with the correct octant for both T2HK and ICAL. However for $\theta_{23}$ in the higher octant, the $\chi^2$ minimum occurs at the wrong octant for ICAL. But when ICAL is added with T2HK, the minimum of the combined $\chi^2$ occurs at the correct octant. Therefore, one can say that for $\theta_{23}$ in the higher octant, T2HK resolves the octant degeneracy of ICAL to improve the mass ordering sensitivity. However, this effect is very small as the ICAL $\chi^2$ with respect to $\theta_{23}$ (test) is almost flat.

\section{Summary and conclusions}
\label{sum}

In this paper we have studied the capability of ICAL, JUNO and T2HK experiments to measure neutrino mass ordering. Though studies on synergy in neutrino oscillation experiments have been performed in the past, in this work we took the opportunity to study the synergy between future accelerator, future atmospheric and future reactor neutrino experiment T2HK, ICAL and JUNO, to determine the neutrino mass ordering. We expounded the synergy in mass ordering determination at these experiments. Mass ordering sensitivity comes from the difference in the oscillation probabilities between normal and inverted ordering. The sensitivity to mass ordering is calculated by simulating the data for an assumed true mass ordering, say normal, and fitted with the wrong mass ordering (say) inverted. In the fit we allow $\delta_{\rm CP}$, $\theta_{23}$ and $|\Delta m_{31}^2|$ to vary within their allowed ranges and the $\chi^2$ is minimized over this entire grid. In particular, we showed that the effect of minimization of the $\chi^2$ over $|\Delta m_{31}^2|$ leads to synergy between the experiments. The best-fit $|\Delta m_{31}^2|$ for which the mass ordering $\chi^2$ is minimized depends on the oscillation channel, neutrino energy, as well as matter effects.  Since T2HK, ICAL and JUNO have different specifications with respect to these, they have synergy with respect to neutrino mass ordering. For this reason, the  same true value of $\Delta m^2_{31}$ leads to different values of $\Delta m^2_{31}$ (fit) for the $\chi^2$ minimum. Therefore, when different experiments are added, the $\chi^2$ minimum of the added $\chi^2$ occurs at a different value of $\Delta m^2_{31}$ leading to an enhanced mass ordering sensitivity. 

We presented results for joint analyses of ICAL+JUNO, ICAL+T2HK and finally ICAL+JUNO+T2HK. We showed that for ICAL+JUNO we can expect about 14\% increase in sensitivity due to synergy as compared to a simple sum of the individual $\chi^2$. However, the enhancement in the case of ICAL+T2HK was not seen to be significant. Finally, when we combined all three ICAL+JUNO+T2HK, we found a staggering increase in the sensitivity. For $\delta_{\rm CP}=0^\circ$ ( $\delta_{\rm CP}=-90^\circ$) this increase was shown to be 331\% (183\%). The ICAL sensitivity has a rather flat dependence on $\Delta m^2_{31}$, while JUNO and T2HK senitivity have a very sharp $\Delta m^2_{31}$ dependence. As a result adding JUNO and T2HK brings the most spectacular increase in the sensitivity. Nonetheless ICAL does play a role and brings additional enhancement to the sensitivity. 

We studied the effect of JUNO energy resolution on the mass ordering sensitivity. We showed that JUNO mass ordering $\chi^2$ would plummet from about 10 to less than 1 if the energy resolution was to worsen from 3$\%/\sqrt{E(MeV)}$ to 5$\%/\sqrt{E(MeV)}$. However, when performing a combined analysis, the synergy between experiments stabilizes the $\chi^2$ and we would have an assured determination  of mass ordering at greater than $7\sigma$  significance even if JUNO has a resolution of 5$\%/\sqrt{E(MeV)}$ and T2HK $\delta_{\rm CP} = 0^\circ$. We also studied the impact of ICAL run-time on mass ordering. We showed that more than $5\sigma$ sensitivity would be easily achievable from different combinations of these experiments and different run times for ICAL. We showed that if ICAL run-time is increases to 20 years, then the combined $\chi^2$ for ICAL+JUNO+T2HK reaches 100 (144) for $\delta_{\rm CP} = 0^\circ (-90^\circ)$. Finally, we studied the impact of $\theta_{23}$ on the mass ordering sensitivity and the role of doing the combined analysis. We also show that if $\theta_{23}$ belongs in the upper octant, then $\chi^2$ minimum for ICAL appears in the wrong octant and the addition of T2HK data resolves this octant degeneracy. 

In summary, we showed that ICAL+JUNO+T2HK is a powerful combination of experiments which can establish the true nature of the mass ordering with a significant confidence level irrespective  of true values of $\delta_{\rm CP}$ and energy resolution of JUNO.

\section*{Acknowledgements}
This work is performed by the members of the INO-ICAL collaboration. We thank to the members of the INO-ICAL collaboration for their valuable comments
and constructive inputs. We thank A. Dighe, S. Goswami for useful discussion and valuable comments during the INO analysis meetings. The authors also sincerely thank the ICAL internal referees, Amol Dighe and D. Indumathi for their careful reading of the manuscript and for providing useful suggestions. Authors extend sincere thanks to Suprabh Prakash for providing the .glb file for JUNO and for useful discussions regarding the JUNO experiment. The HRI cluster computing facil-
ity (http://cluster.hri.res.in) is gratefully acknowledged. MG acknowledges Ramanujan Fellowship of SERB, Govt. of India, through grant no: RJF/2020/000082. This project has received funding/support from the European Union’s Horizon 2020 research and innovation programme under the Marie Skłodowska -Curie grant agreement No 860881-HIDDeN.

\bibliographystyle{JHEP}
\bibliography{hierarchy_INO_JUNO_T2HK}
  
\end{document}